\documentclass[aps, pra,twocolumn, showpacs,superscriptaddress,10pt,floatfix]{revtex4-2}

\usepackage{graphicx}
\usepackage{dcolumn}
\usepackage{bm}
\usepackage{hyperref}
\usepackage{xcolor}
\usepackage{mathrsfs}
\usepackage{physics}
\usepackage[page]{appendix}
\usepackage{tikz}
\usepackage{bbold}

\definecolor{persiangreen}{rgb}{0.0, 0.65, 0.58}

\newcommand{\JO}[1]{{\color{magenta}{#1}}} 

\newcommand{\pcsadd}{Center for Theoretical Physics of Complex Systems, Institute for Basic Science (IBS), Daejeon - 34126, Korea}
\newcommand{\ustadd}{Basic Science Program, Korea University of Science and Technology (UST), Daejeon - 34113, Korea}

\begin{document}

\title{Artificial intelligence discovery of a charging protocol in a micromaser quantum battery}

\author{Carla Rodríguez}
\email{carla.rodriguez@mpl.mpg.de}
\affiliation{Max Planck Institute for the Science of Light, Erlangen, Germany}
\affiliation{Optics Group, Physics department, Universitat Autònoma de Barcelona, Barcelona, Spain}
\author{Dario Rosa}
\email{dario\_rosa@ibs.re.kr}
\affiliation{\pcsadd}
\affiliation{\ustadd}
\author{Jan Olle}
\email{jan.olle@mpl.mpg.de}
\affiliation{Max Planck Institute for the Science of Light, Erlangen, Germany}

\date{\today}

\begin{abstract}
We propose a gradient-based general computational framework for optimizing model-dependent parameters in quantum batteries (QB). We apply this method to two different charging scenarios in the micromaser QB and we discover a charging protocol for stabilizing the battery in upper-laying Hilbert space chambers in a controlled and automatic way. This protocol is found to be stable and robust, and it leads to an improved charging efficiency in micromaser QBs. Moreover, our optimization framework is highly versatile and efficient, holding great promise for the advancement of QB technologies at all scales.
\end{abstract}

\maketitle

\section{Introduction}
\label{sec:intro}

The world is currently immersed in the "second quantum revolution" and its associated quantum technologies~\cite{dowling2003quantum, preskill_quantum, Flamini_2019, georgescu_quantum_simulation}, as evidenced by the increasing investment from governments and industries on these technologies, which promise to revolutionize our society in the near future~\cite{Acin2018}. 
Put in simple terms, quantum technologies are devices that leverage inherently quantum phenomena like coherence and entanglement to solve tasks that, in some cases, a classical device would never be able to solve in a reasonable amount of time.
The possibility of using quantum mechanical effects to outperform classical machines has been dubbed \textit{quantum advantage}, and the most famous examples are the recently achieved milestones in quantum information processing~\cite{Arute2019, science_quantum_advantage, Madsen2022}.
Simultaneously, the trend of miniaturization in technology, with devices operating at the nanoscale, has resulted in an extension of the traditional thermodynamic concepts such as work and heat to account for quantum mechanical effects, leading to the emergence of the field of \textit{quantum thermodynamics}, see, for example, \cite{vinjanampathy2016quantum, binder2019thermodynamics, deffner2019quantum} for few reviews on the topic.
Intersecting between these two active research topics, the concept of quantum battery has emerged and flourished \cite{campaioli2018quantum, bhattacharjee2021quantum}, both as a necessary step to provide energy to nanodevices as well as a suitable sub-field of quantum technologies where quantum effects can provide various sources of advantages, even at larger scales.

As their name suggests, quantum batteries are quantum mechanical systems that can be used to store energy, which can then be released as needed.
This energy is stored in a highly excited state (with respect to its Hamiltonian) and released by driving the system to a lower energy state.
The first work on the subject was by Alicki and Fannes \cite{alicki_fannes}, after which subsequent research works extended in several different directions. These include the proposal of explicit theoretical models of quantum batteries \cite{liu_loss-free_2019, rossini_many-body_2019,le_spin-chain_2018,rossini_quantum_2020,ferraro_high-power_2018, seah2021quantum, salvia2022quantum}
and the analysis of the figures of merit to address the efficiency of such devices, such as the charging power \cite{campaioli_enhancing_2017, PhysRevResearch.2.023113}, the maximum amount of energy which can be unitarily extracted \cite{delmonte_twophotonQB, Allahverdyan_2004,safranek_work_extraction}, the effect of noise and imperfections \cite{Shnirman_2002, ciccarello_collision_models, Carrega_2020, caravelli2020random}, non-markovian effects \cite{Kamin_2020, morrone2022charging}, stability and fluctuations properties \cite{Rosa2020} or the necessary conditions to achieve a \emph{quantum charging advantage} \cite{PhysRevLett.128.140501, konar2022quantum, mondal2022periodically}.

At the same time, although at a slower pace, the first experimental realizations have been proposed and studied \cite{quach_superabsorption_2022, gemme_ibm_2022}.
On general grounds, a promising platform for the development of quantum batteries is quantum electrodynamics, where the interaction between light and matter can be currently manipulated with high precision \cite{forn-diaz_ultrastrong_2019}.
A well-established electrodynamical setup -- studied both theoretically and experimentally -- is the micromaser \cite{filipowicz_quantum_1986, PRL_Maser, slosser_harmonic_1989, slosser_generation_1990, slosser_tangent_1990, LeKien1993}.  In a micromaser, a stream of two-level systems (qubits) interacts sequentially with the electromagnetic field in a cavity, modelled as a quantum harmonic oscillator.
Indeed, this system has recently been proposed as an excellent model of a quantum battery \cite{shaghaghi_micromasers_2022, shaghaghi_lossy_2022} and can be described by the celebrated Jaynes-Cummings model \cite{jaynes_comparison_1963}. 

Optimization is a crucial aspect of many scientific and engineering problems \cite{ML_trends, lecun2015deep, Dawid_modern_applications_ML, highbias_lowvariance}, and Artificial Intelligence (AI) methods have shown to be particularly powerful in these regards. Within the field of quantum technologies, AI has demonstrated remarkable versatility and efficacy through a diverse array of methods and applications, \cite{carleo_machine_2019, krenn_artificial_2023, Dunjko_2018}. 
Optimal control methods for quantum batteries have been explored \cite{optimal_control_methods_andolina, optimal_quantum_control}. However, with the only very recent exception of \cite{erdman_reinforcement_2022}, the literature on using AI methods to optimize aspects of QBs is still missing.

In this work, we present a general AI-based computational framework that optimizes the performance of quantum batteries. To exemplify it, we choose the micromaser as a QB model and optimize model-dependent parameters using gradient descent. However, this approach can be easily adapted to accommodate a variety of models and a variety of different figures of merit quite straightforwardly.

In our approach, we have deliberately chosen an optimization algorithm that is as simple as possible. By doing this, we aim to circumvent the potential obfuscation of scientific understanding that can occur when utilizing complex, black-box AI algorithms, without compromising its performance. This philosophy is inspired by recent works on digital discoveries of quantum optical experimental setups \cite{arlt_digital_2022, ruiz-gonzalez_digital_2022}. 

As a first practical result of our approach, we discover a charging protocol for micromaser QBs that allows to substantially increase its energy storage while keeping the battery stable. This result shows concretely the power of simplicity in gaining insight into the problem at hand.

The paper is organized as follows: In Section \ref{sec:optimization}, we introduce and describe our optimization framework, including the use of computational graphs, the overview of the micromaser QB model and the specific cases of interest we have studied for optimization and a discussion of the loss function used in our analysis. In Section \ref{sec:results}, we present the numerical results. Finally, in Section \ref{sec:conclusions}, we provide our conclusions and outline the potential future directions of this work.

\section{Optimization framework}
\label{sec:optimization}

In this section, we introduce an optimization framework --- suitable for general models of quantum batteries --- that optimizes the charging process using gradient descent. The gradient descent method is a widely-used optimization technique for finding the minimum of a differentiable function of interest, typically called a loss function $\mathcal{L}$. One of the main reasons for its popularity is its effectiveness in finding good solutions to a wide range of optimization problems. In addition, it is relatively simple to implement, making it a useful tool for practitioners in various fields \JO{\cite{Ruder16_review}}. Given an initial set of model parameters, $\Vec{p}$, the algorithm iteratively adjusts $\Vec{p}$ in the direction of the negative gradient of the loss function with respect to the model parameters as follows:
\begin{equation}
    \Vec{p}~\rightarrow~\vec{p} - \eta \frac{\partial \mathcal{L}}{\partial \vec{p}}~,
\end{equation}
where $\eta$ is the hyperparameter known as the learning rate, which controls the step size of the update (i.e., the fraction of the gradient to be subtracted from the original parameters). The optimization process terminates when the loss function reaches an acceptable minimum value set by some tolerance.

\begin{figure}[h]
  \centering
  \includegraphics[width=0.5\textwidth]{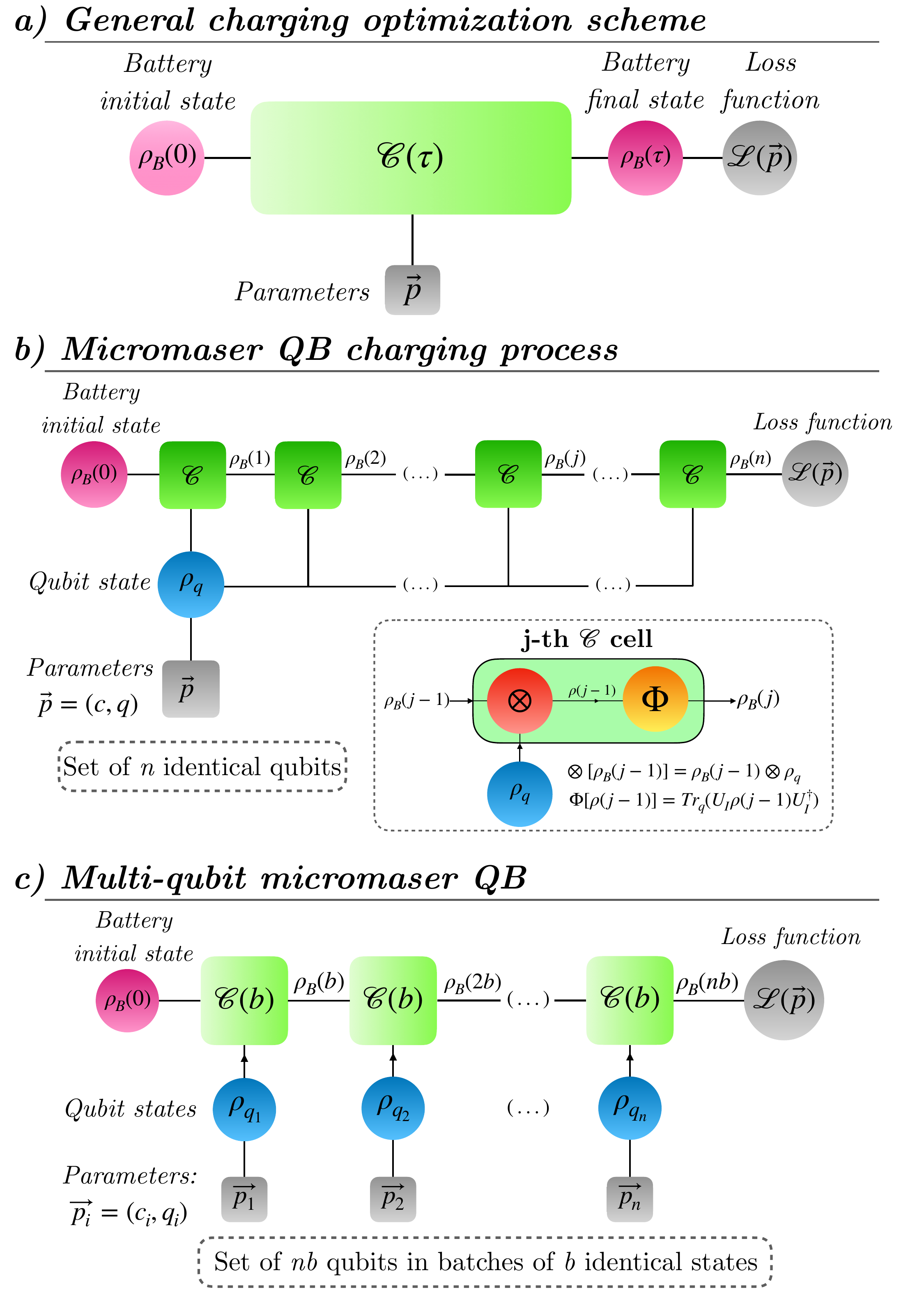}
  \caption{(a) General representation of the optimization of the charging process for a quantum battery using a computational graph. The battery is initialized in state $\rho_B(0)$ and is charged through cell $\mathcal{C(\tau)}$, during time $\tau$, which is represented, in general, by a completely-positive and trace-preserving (CPTP) map. This process results in the final battery state $\rho_B(\tau)$. The loss function $\mathcal{L}(\Vec{p})$, which is dependent on the model parameters $\Vec{p}$, is calculated at the end of the charging process and it is used for updating the parameters in the next iteration of the algorithm. Panels (b) and (c) show the computational graph representing the optimization of the charging process of a micromaser QB. Whereas (b) shows the system with $n$ identical qubits prepared in state $\rho_q$, characterized by parameters $\Vec{p}$, panel (c) shows the system with $n\cdot b$ qubits in batches of $b$ identical states $\rho_{q_i}$. Each charging cell $\mathcal{C}$ in panel (b) corresponds to a single cavity-qubit interaction and consists of the operations $\bigotimes$ and $\Phi$, as indicated inside the dashed box. In panel (c), each charging cell $\mathcal{C}(b)$ represents the sequential interaction of a batch of $b$ identical qubits, as seen in panel (b).}
  \label{fig:General-model}
\end{figure}

The proposed optimization approach is illustrated in Fig. \ref{fig:General-model}a. It has several key advantages. First, it is versatile and can be applied to a wide range of quantum battery models (including those involving large-scale systems) described by a set of parameters $\Vec{p}$, as represented by the cell $\mathcal{C}$ in Fig. \ref{fig:General-model}a. Second, it is efficient 
and able to find the optimal solution to a problem in an automated manner given a loss function, $\mathcal{L}(\Vec{p})$, that depends on the model parameters $\Vec{p}$ either implicitly or explicitly. Overall, this optimization framework offers a promising solution for optimizing many different aspects of quantum battery models.

\subsection{Optimization of a micromaser QB}
\label{sec:optimization_of_a_micromaser_QB}

The micromaser \cite{shaghaghi_micromasers_2022} charging process consists of the sequential interaction of a single mode of an electromagnetic field in a cavity with a stream of two-level systems (qubits) prepared in a state which can be written as
\begin{equation}
    \rho_q = q\ket{g}\bra{g} + (1-q)\ket{e}\bra{e} + c\sqrt{q(1-q)}(\ket{g}\bra{e} + \ket{e}\bra{g}), 
\end{equation}
where $q$ is the degree of population inversion, $c$ the coherency, and $\ket{g}$ and $\ket{e}$ correspond to the ground and excited state of the qubit, respectively.

In the qubit-cavity interaction picture, the evolution of the system is described by the Hamiltonian \cite{jaynes_comparison_1963}, 
\begin{equation}
   \hat{H}_{I}= g(\hat{a} \hat{\sigma}_{+} + \hat{a}^\dag\hat{\sigma}_{-} + e^{2i\omega t}\hat{a}^\dag \hat{\sigma}_{+} + e^{-2i\omega t}\hat{a} \hat{\sigma}_{-}),
    \label{eq:H_I}
\end{equation}
where $g$ represents the coupling constant between the qubit and the field, and with units chosen such that $\hbar = 1$. The lowering and raising operators for the qubit are denoted by $\hat{\sigma}_{-}$ and $\hat{\sigma}_{+}$, respectively; $\hat{a}$ and $\hat{a}^\dag$ are the annihilation/creation operators for the field, respectively. We are examining the resonant case, where the frequencies of the qubit and field are equal \cite{meystre_elements_1991} and here denoted by $\omega$.

To optimize the charging speed of a quantum battery, the ultra-strong coupling regime, $0.1 < \frac{g}{\omega} < 1$ is the most interesting regime to consider. This is because larger values of $g$ are associated with larger values of the charging power during the charging process. This relationship between the charging power and $g$ has been shown numerically in \cite{shaghaghi_micromasers_2022}. Heuristically, one can understand this relationship because $g$ controls the amount of energy that the qubit is able to transfer to the micromaser during the interaction time. Hence, larger values of $g$ are associated with more energy transferred by the qubit and, consequently, larger values of the power.
Unfortunately, in this regime the counter-rotating terms (the last two terms in Eq.~\eqref{eq:H_I}) are relevant and they make the dynamics much more complicated to describe.
In particular, even the choice of which Hamiltonian must be used in this regime is a matter of active debate \cite{DiStefano2019, Stokes2019}.
In this paper, to avoid these complications, we will make use of the approach introduced in \cite{huang_ultrastrong_2020}, which shows that the counter-rotating terms can be neglected by performing a simultaneous frequency modulation of the qubit and the field, and the resulting dynamics is described by the Jaynes-Cummings unitary operator~\cite{jaynes_comparison_1963}, 
\begin{equation}
    \hat{U}_I(g) = e^{-i\tau g(\hat{a} \hat{\sigma}_{+} + \hat{a}^\dag\hat{\sigma}_{-})},
    \label{eq:U_I}
\end{equation}
where $\tau=1$ can be fixed without affecting the validity of the approach, \footnote{Here we want to stress that the proposal of \cite{huang_ultrastrong_2020}, while perfectly correct in theory, could be hard to implement in an experimental setup.
For this reason, a thorough analysis of the results of this paper in the presence of counter-rotating terms represents an interesting open problem. However, it is important also to observe that taking the value of $g$ below the critical value of $0.1$, \textit{i.e.} within the regime of validity of the Jaynes-Cummings approximation, does alter the results only \textit{quantitatively} and not qualitatively.}. 

Moreover, the coherence parameter $c$ can be set to be real by applying a rotation along the $z$ axis which does not alter the time evolution operator, $\hat{U}_I(g)$.

The battery charging process, as depicted in Fig. \ref{fig:General-model}b, is initialized by the cavity in its ground state $\rho_B(0) = \ket{0}\bra{0}$. The joint qubit-cavity system is obtained by the tensor product
\begin{equation}
    \rho(k) = \otimes[\rho_B(k)] \equiv \rho_B(k) \otimes \rho_q,
    \label{eq:tensor}
\end{equation}
being $\rho_B(k)$ the battery state after $k$ interactions, $k=0, \dots, n$. The time evolution can be recursively written as
\begin{equation}
    \rho_B(k+1)  = \Phi[\rho(k)] \equiv \text{Tr}_q(\hat{U}_I(g)\rho(k)\hat{U}_I^\dag(g)).
\end{equation}
where $\rho(k)$ corresponds to the system state resulting from the previous Eq.~\eqref{eq:tensor}, and $\text{Tr}_q$ is the trace over qubit degrees of freedom. This outputs the battery state $\rho_B(k+1)$.

The charging process can be represented computationally by iteratively applying the operations $\bigotimes$ and $\Phi$ as shown in Fig. \ref{fig:General-model}b. The system consists of an initialized battery in state $\rho_B(0)$ and a set of $n$ identical qubits prepared in state $\rho_q$, characterized by the parameters $\vec{p} = (c, q)$. The interaction between the cavity and the qubit is represented by the cell $\mathcal{C}$, where the charging process occurs. The resulting output battery state, $\rho_B(1)$, is injected into the next charging cell, $\mathcal{C}$, where it interacts with a new qubit. This process is repeated for $n$ interactions, resulting in the final battery state $\rho_B(n)$. In the optimization framework, the loss function $\mathcal{L}(\Vec{p})$ is evaluated after each charging cycle of $n$ identical qubits. This function, which depends on the qubit state parameters $\Vec{p}=(c,q)$, is minimized by repeatedly updating the parameters after each complete charging cycle, using a gradient-descent optimization algorithm. 

This process can be generalized to a multi-qubit scheme as shown in Fig.~\ref{fig:General-model}c. In this scheme, qubits prepared in different states are processed in different batches of $b$ identical qubits. Each batch interacts with the cavity in the same way as described in Fig.~\ref{fig:General-model}b. Similarly, the charging process is repeated for $n$ batches, resulting in the final battery state $\rho_B(nb)$. In this context, the parameters to be optimized are no longer a single pair $(c,q)$, but $n$ pairs of parameters $(c,q)$ for each batch of $b$ identical qubits.

\begin{figure}[h]
  \centering
  \includegraphics[width=0.48\textwidth]{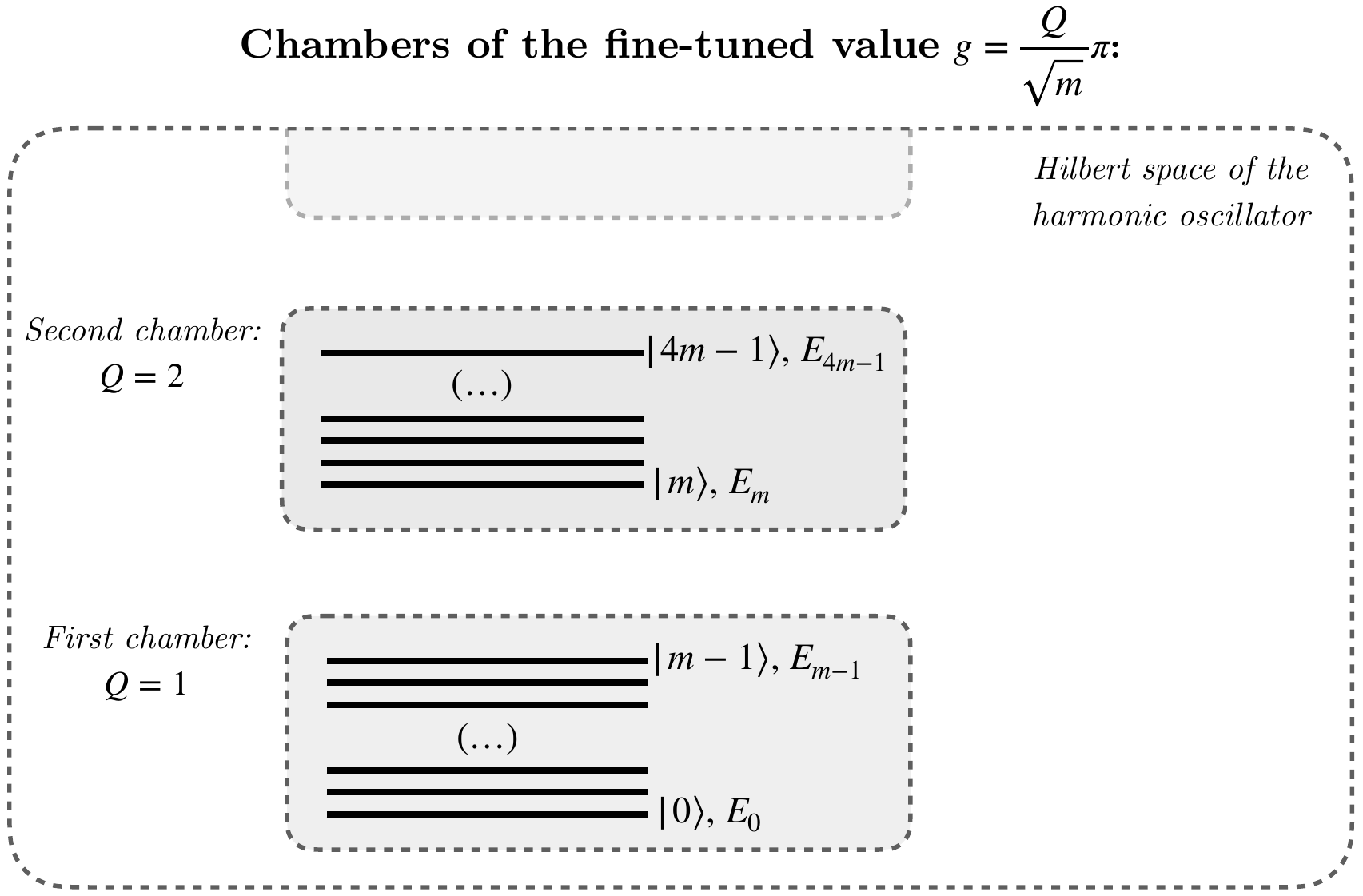}
  \caption{Sketch of trapping chambers for the fine-tuned value $g=\frac{Q}{\sqrt{m}}\pi$. The first chamber is defined by $Q=1$ and contains the states $\ket{n}$ such that $n<m$. The second chamber is defined by $Q=2$ and contains the states $\ket{n}$ satisfying $m<n<4m$. For the purpose of illustration, we only depict the lowest-lying two chambers. By unitary time evolution, the initial chamber cannot be exited, effectively \textit{trapping} the system there. $E_{m-1} = m-1$ and $E_{4m-1} = 4m-1$ are the energies of the highest excited states corresponding to the first and the second chambers, respectively.}
  \label{fig:Chambers}
\end{figure}

One of the most appealing features of a micromaser QB is the existence of dynamically separated trapping chambers \cite{filipowicz_quantum_1986, slosser_harmonic_1989, slosser_tangent_1990, slosser_generation_1990}, whose relevance for energy storage purposes has been analyzed in details in \cite{shaghaghi_micromasers_2022, shaghaghi_lossy_2022}. 
The existence of the trapping chambers is analytically demonstrated in the literature \cite{slosser_harmonic_1989, slosser_tangent_1990, filipowicz_quantum_1986} for the particular case of $c=1$ (qubits are fully coherent, i.e. they are in a \emph{pure} state) and $g$ being fine-tuned to
\begin{equation}
    g = \frac{Q}{\sqrt{m}}\pi, \qquad  Q,m\in \mathbb{N},
    \label{eq:g_definition}
\end{equation}
where $Q$ and $\sqrt{m}$ are integer which do not share any common integer divisors. When these conditions are met, the harmonic oscillator Hilbert space dynamically separates the states $\ket{n}$ with $n < m$ from the remainder of the Hilbert space. By doing a simultaneous redefinition of $\tilde{Q}=2Q$ and $\tilde{m}=4m$, the same argument leads to a dynamical separation of states $|n\rangle, \; m < n < 4m$ from the rest. Repeating the same argument \textit{ad infinitum} leads to a fragmentation of the Hilbert space in dynamically separated chambers, meaning that one cannot exit the initial chamber by time evolution \cite{shaghaghi_micromasers_2022}. This is illustrated in Fig. \ref{fig:Chambers}.

When $g$ is fine-tuned according to Eq.~\eqref{eq:g_definition}, two charging protocols are known to exist with an analytic description \cite{shaghaghi_lossy_2022}. The first one is the \textit{coherent charging protocol}, occurring when the qubits are prepared in a coherent superposition ($c=1$). In this case, after it has completely charged, the battery reaches a steady state which is pure~\footnote{Given an arbitrary density matrix $\rho$, its purity is defined as $\mathcal{P}\equiv Tr(\rho^2)$, being equal to one for pure states.} and sustained by the inherent dynamics of the battery. The populations of such steady state satisfy the following recursive relation
\begin{equation}
    \rho_{B}^{(n,n)} = \frac{1-q}{q} \cot^2\left(\frac{\pi}{2\sqrt{m}}\sqrt{n}\right)\rho_B^{(n-1,n-1)}, \label{eq:cotangent_states}
\end{equation}
where $n<m$. On the other hand, incoming qubits prepared in an incoherent mixture ($c=0$) define the \textit{incoherent charging protocol}. For $q = 0$, the battery steady state is pure and given by the number state $\ket{m-1}$. However, as $q$ becomes nonzero, its purity decreases as $\mathcal{P} \approx 1 - 2q$ to first order in $q$.

Importantly, while creating a pure steady state, the \textit{coherent} charging procedure results in a reduction on the total energy stored in the battery as compared to the \textit{incoherent} charging case, making purity an energetically costly quantity. This can be computed from Eq.~\eqref{eq:cotangent_states}.

For practical applications, the regime in which $g$ is not fine-tuned must be considered, and it was also the subject of \cite{shaghaghi_micromasers_2022,shaghaghi_lossy_2022}. This can be achieved by introducing a parameter $\epsilon$ as
\begin{equation}
    g = \frac{Q}{\sqrt{m + \epsilon}}\pi, \qquad  Q,m\in \mathbb{N}, \; -0.5 < \epsilon \leq 0.5~.
    \label{eq:g_approx}
\end{equation} 
A subsequent numerical analysis showed that, by allowing small perturbations on either the fine-tuned value of $g$ or $q \neq 0$, the \textit{incoherent} charging protocol just described deviates away from the fine-tuned steady state $\ket{m-1}$ and instead results in an indefinite energy growth and thus a non-existent steady state. On the other hand, for all values of $q$ and even in the presence of non fine-tuned values of $g$, the \textit{coherent} charging protocol allows to reach an effectively pure steady state with the same properties as in the fine-tuned regime (the steady state just described is, plausibly just meta-stable, but its lifetime has been tested to be very long by the extensive numerical simulations performed in \cite{shaghaghi_micromasers_2022}).

Interestingly, Eqs. \eqref{eq:g_definition} and \eqref{eq:g_approx} and the consequent structure of the chambers, allow us to obtain an \textit{inverse} relationship between the value of $g$ and the energy (and, to a large extent, the ergotropy as well) of the steady state: since large values of $g$ are associated to smaller values of $m$, we see that the energy of a chamber is reduced by increasing $g$.

\subsection{Loss function}
\label{sec:loss_function}

The selection of an appropriate loss function is perhaps the most delicate step in the context of posing an optimization problem. This is due to the loss function serving as the basis for determining how to modify the parameters of the model in order to improve its performance. If an inappropriate loss function is chosen, it may be challenging or even impossible to find a satisfactory solution, as the optimization process will be attempting to minimize a loss function that is not well-suited to the problem at hand.

In the context of quantum batteries, a quantity of interest to be maximized is the ergotropy, $\mathcal{W}$, defined as the maximal amount of work that can be extracted from the battery with a unitary operation. It can be computed according to \cite{rossini_many-body_2019, tirone_quantum_2022},
\begin{equation}
    \mathcal{W} \equiv E - E_{\rm{Passive}} = \text{Tr}(H_F\rho_B) - \sum_k r_k\varepsilon_k,
    \label{eq:ergotropy_definition}
\end{equation}
where $E_{\mathrm{Passive}}$ denotes the energy of the \textit{passive} state~\footnote{A quantum state is said to be passive when no further energy can be unitarily extracted from it.} associated to the battery state $\rho_B$. It is computed with $r_k$ and $\varepsilon_k$, which are the eigenvalues of the battery density matrix $\rho_B$, and the field Hamiltonian, $H_F = \omega \hat{a}^\dag \hat{a}$, respectively.
It is easy to show that when the battery state is \textit{pure}, i.e. when it can be written as $\rho_B = \ket{\psi} \bra{\psi}$ for an appropriate state $\ket{\psi}$, the associated energy $E_\mathrm{Passive}$ turns out to be the Hamiltonian ground state energy $\epsilon_0$ (which we can always set to zero without loss of generality).
Thus, in this case it is always possible to bring the battery state to the ground state via a unitary operation, and the ergotropy coincides with the energy of that state.
On the other hand, if the battery is in a \textit{mixed} state, it obviously cannot evolve unitarily to the Hamiltonian ground state.
In such a case, the maximum amount of energy that can be extracted is computed by $\mathcal{W}$ as in Eq.~\eqref{eq:ergotropy_definition} and it satisfies $\mathcal{W} < E$. It then follows that a way to increase the ergotropy while keeping constant the energy of the battery state, $E$, is by increasing its degree of purity $\mathcal{P}$.
When considering the micromaser battery, an essentially pure state can be reached when the incoming qubits are fully coherent with $c = 1$.
However, the value of $\mathcal{W}$ can grow by increasing the energy $E$ of the battery state.
As a result, maximizing the ergotropy effectively means finding a good compromise between a high value of the battery energy and a sufficiently large value of its purity.

A specific feature of the micromaser QB that we have reviewed in Sec.~\ref{sec:optimization_of_a_micromaser_QB} is the existence of very long-lived metastable steady states. These are appealing because they are a built-in feature of the model that avoids the battery absorbing an unbounded amount of energy in an uncontrolled way. Steering away from these unwanted states must also be incorporated in the definition of the loss function, and this can be achieved by including a penalty term for large values of the time-derivative of the energy of the battery state at the terminal stages of the charging process.

The final two considerations that must be taken are related to specific values of the parameters $c,q$ themselves. Related to the former, maintaining high values of $c$ in qubit states is a very challenging experimental task, while qubits carry no energy when $q=1$ since they are prepared in their ground state. Thus, the loss function must also penalize large values of both $c$ and $q$.

Summarizing, our task is to construct a loss function that maximizes the ergotropy while penalizing both large values of $c$ and $q$ and short-lived metastable steady-states. With these requirements in mind, we consider the loss function for a charging process with $n$ identical qubits to be
\begin{equation}
    \mathcal{L} = - \frac{\mathcal{W}_n \cdot (1-c) \cdot (1-q)}{\mathcal{W}_n + (1-c) + (1-q)} + \lambda \sum_{k \in \eta} \left|\frac{\Delta E_k}{\Delta k}\right|~,
    \label{eq:loss_function_single}
\end{equation}
where $\mathcal{W}_n$ is taken in units $\omega=1$ and where we have introduced two hyperparameters $\eta$ and $\lambda$; $\eta$ is a subset of qubits, typically a fraction of the last ones, and $\lambda$ is a positive number that penalizes unstable steady-states. When the qubits that interact with the cavity are not identical, Eq.~\eqref{eq:loss_function_single} gets modified by replacing $c,q$ by their mean value over the set of qubits.

The first term in Eq.~\eqref{eq:loss_function_single} looks counterintuitive at first, so let us elaborate further on this point. Its structure stems from wanting to simultaneously optimize three quantities, $\mathcal{W}_n,~ c,$ and $q$, which have very different magnitudes, $\mathcal{W}_n \gg c,q$. The first term is a convenient way to normalize these three terms, guaranteeing that each of them has an equiparable weight during the optimization. An alternative would be having a separate term for each of these three quantities, but this would introduce two extra hyperparameters to be varied. As a result, the computational cost of the optimization process would have also increased, and this is why we have discarded this path for now.

Another point to address is the fact that large values of $q$ are already obviously penalized when maximizing the ergotropy $\mathcal{W}$, so it may seem that including $q$ explicitly in the loss function is not necessary. We have experimented with such a loss function, and we have found it to have a landscape that in some cases benefits reaching $q=1$, especially in the case where not all the qubits are identical.

Finally, we have used the energy and not the ergotropy in the second term of Eq.~\eqref{eq:loss_function_single} since the behaviour of their variation is extremely similar and the ergotropy is particularly computationally costly since it requires performing a diagonalization.

\section{Results}
\label{sec:results}

In this section we present the optimization results conducted for the micromaser charging process using (i) a set of $n$ identical qubits and (ii) a set of two batches consisting of $b$ identical qubits in each batch. While setup (i) serves as a sanity benchmark for our methods, in (ii) we discover a charging protocol that is able to stabilize the battery to a state laying in the second chamber. We have written the physics simulation and the implementation of the optimization algorithm using Python's NumPy and SciPy packages \cite{numpy, scipy}. More specifically, the optimization is conducted using the 'L-BFGS-B' solver from SciPy, which is a method that computes gradients numerically and allows for parameters to be bounded. The optimization is initialized using a randomized pair of $(c,q)$. From now on we will work in units $\omega = 1$ and truncate the infinite-dimensional quantum harmonic oscillator Hilbert space to a maximal number of modes of $N_c=120$ and $N_c=150$ in the numerical simulations of the stream of identical qubits and different qubit batches with identical states within, respectively. We have checked that the results are insensitive to this truncations, i.e. the truncations are taken at levels which are never populated during the dynamical evolution.

\subsection{Stream of identical qubits}
\label{sec:results_identical}

We conduct the optimization of the micromaser charging process described in Fig. \ref{fig:General-model}b, using a stream of $n=1000$ identical qubits interacting with the battery. The coupling constant is set to a non fine-tuned value of $g=\pi/\sqrt{15.6} \approx 0.795$. We utilize the loss function $\mathcal{L}$ in Eq.~\eqref{eq:loss_function_single} with $\eta$ fixed to be the last $20\%$ of qubits and consider several values of the hyperparameter $\lambda$. For each case, we find a set of optimal values of the pair $(c,q)$ that minimize $\mathcal{L}$. Our results show that for $\lambda = 1$, the optimal $(c,q)$ values are approximately $(0.310, 0.145)$; for $\lambda = 10$, $(c,q) \approx (0.390, 0.168)$; for $\lambda = 100$, $(c,q) \approx (0.470, 0.183)$; and for $\lambda = 1000$, $(c,q) \approx (0.553, 0.190)$, from which we see that by increasing $\lambda$, which corresponds to increasing the importance of the stability of the final state, there is a general trend in seeking an enhancement of $c$.

Using the optimized $(c,q)$ parameters, we simulate the time evolution described by Eq.~\eqref{eq:U_I} and calculate the ergotropy $\mathcal{W}$, as given by Eq.~\eqref{eq:ergotropy_definition}, for up to $k=10^5$ collisions to evaluate the stability of the steady state for different values of $\lambda$. The results of this simulation were then compared to those obtained using the closest fine-tuned (FT) value of $g = \pi/\sqrt{16} \approx 0.811$ using the \textit{incoherent} charging protocol. 

\begin{figure*}
  \centering
  \includegraphics[width=\textwidth]{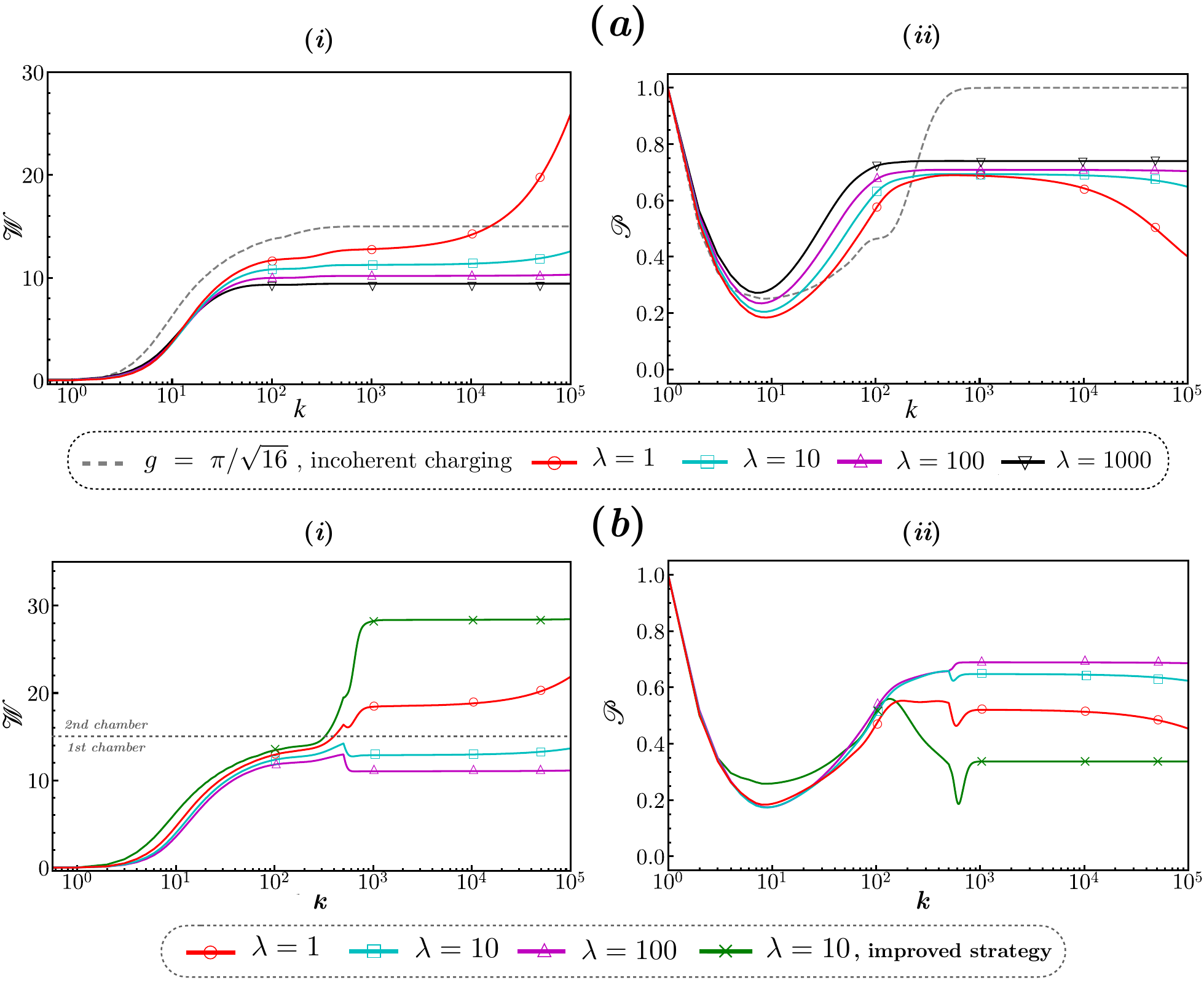}
  \caption{Results for the Ergotropy $\mathcal{W}$ (in units $\omega=1$) and Purity $\mathcal{P}$ of the non fine-tuned value of $g=\pi/\sqrt{15.6}$ as a function of the number of collisions $k$ with parameters obtained from the optimization procedure. Panel (a) shows the results for the charging scheme that uses a stream of identical qubits while (b) depicts the corresponding one for a stream of different qubit batches of identical states within. The optimal parameters $(c,q)$ found in (a) are $(c,q) \approx (0.310, 0.145)$ (red, circle), $(c,q) \approx (0.390, 0.168)$ (cyan, square), $(c,q) \approx (0.470, 0.183)$ (magenta, up-triangle), $(c,q) \approx (0.553, 0.190)$ (black, down-triangle), compared with the maximum ergotropy achieved in the closest fine-tuned value of $g=\pi/\sqrt{16}$ and incoherent charging (dashed). The optimal parameters found for (b) are $\{\Vec{p}_1, \Vec{p_2}\} = \{(c,q)_{\rm{Batch}=1},(c,q)_{\rm{Batch}=2}\}$: $\{(0.150,0.079), (0.361, 0.150)\}$ (red, circle), $\{(0.229,0.123), (0.414, 0.165)\}$ (cyan, square), $\{(0.270, 0.152), (0.486, 0.179)\}$ (magenta, up-triangle), $\{ (0,0), (0.449,0.208) \}$ (green, cross). In all cases, the effect of the hyperparameter $\lambda$ is inversely proportional to ergotropy. It is also observed that a loss in purity results in a gain of ergotropy.}
  \label{fig:single_qubit}
\end{figure*}

The results of this comparison are shown in Fig.~\ref{fig:single_qubit}a(i). We see that the saturation to a steady state occurs more quickly for the non-fine tuned value of $g=\pi/\sqrt{15.6}$. However, these states are metastable. The effect of the hyperparameter $\lambda$ on the system can be seen by comparing the different solid lines in Fig. \ref{fig:single_qubit}a(i): a smaller value of $\lambda$ results in a shorter-lived steady state, but an increase in ergotropy $\mathcal{W}$. The fine-tuned value (dashed line in Fig. \ref{fig:single_qubit}a) achieves a maximal ergotropy of $\mathcal{W}=15$ at $q=0$ and a steady state (number state) that is absolutely stable (although fragile for small perturbations of the parameters), in agreement with the results for the \textit{incoherent} charging protocol discussed in Section \ref{sec:optimization_of_a_micromaser_QB}. Values of $\lambda$ above $10$ (magenta (up-triangle) and black (down-triangle) lines in Fig. \ref{fig:single_qubit}a(i), respectively) show no signs of instability for the time scales that we have simulated. The case $\lambda=10$ (cyan (square) line in Fig. \ref{fig:single_qubit}a(i)) shows an instability after roughly $10^4$ interactions, while there is no sign of stability for the remaining $\lambda=1$ (red, circle) line.
Besides $\lambda=1$, all states achieved ergotropies ranging between $9$ and $11$, which amounts to approximately $65\%$ of the closest fine-tuned maximal ergotropy value.

From this numerical considerations, we learn the importance of the hyperparameter $\lambda$: it allows, in realistic scenarios, to maximize the ergotropy while reaching the lifetime required by the particular application under consideration.

An additional characterization of the charging process can be obtained by computing the purity for the same pairs $c,q$ of optimized parameters that we have shown in Fig.~\ref{fig:single_qubit}a. We show the time evolution of the purity of such states in Fig.~\ref{fig:single_qubit}a(ii), from which it becomes apparent that such gains in ergotropy come at the expense of a loss in purity. In this situation, the mean energy of the battery state increases, suggesting that steady states become more unstable as they approach the upper level of the first chamber.

\begin{figure*}
  \centering
  \includegraphics[width=\textwidth]{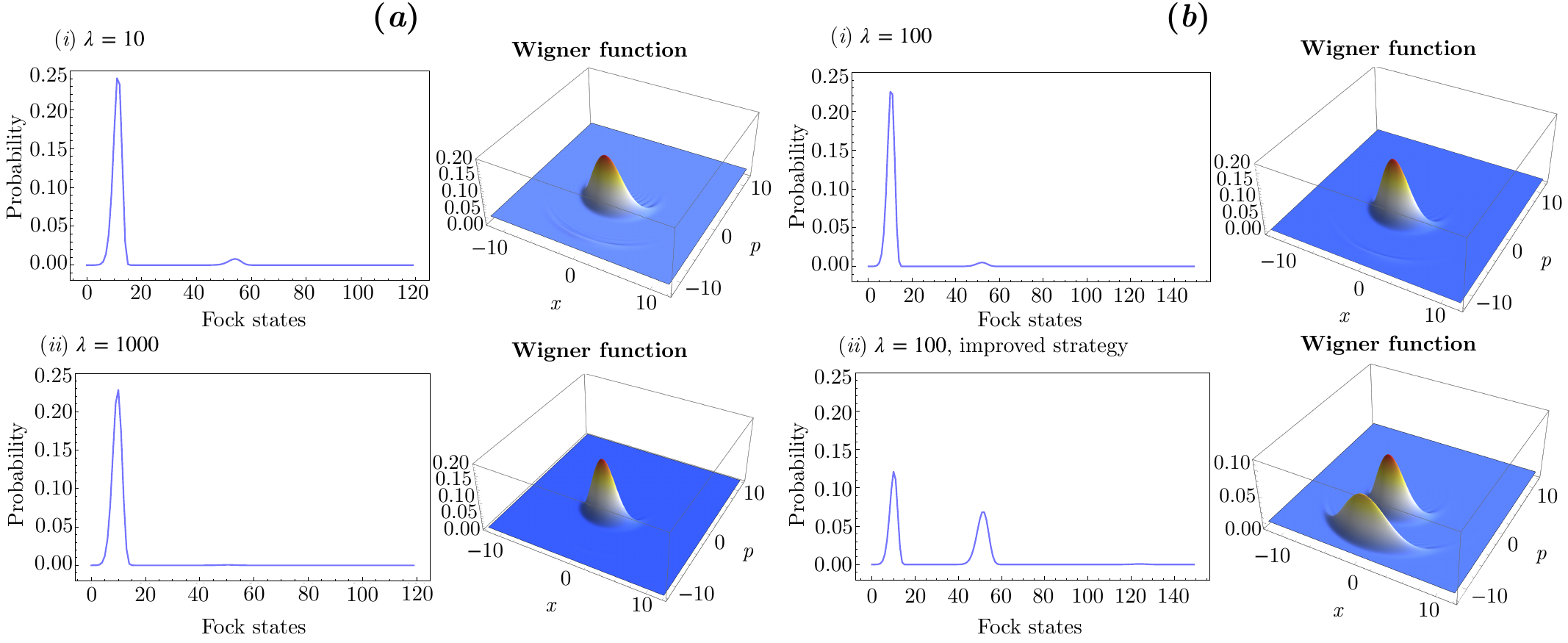}
  \caption{Population distribution of number states $\rho_B^{(n,n)}$ in the steady state of the battery $\rho_B(k)$ after $k=10^5$ collisions with the optimized $c,q$ parameters shown in Fig. \ref{fig:single_qubit} and its associated Wigner function for (a) Charging with a stream of identical qubits and (b) Charging with a stream of different qubit batches of identical states within.
  }
  \label{fig:single_qubit_wigner}
\end{figure*}

The charged battery steady state can be further characterized by computing the population of each number state present in the mixture and its associated quasi-probability Wigner function \cite{wigner, doi:10.1063/1.5046663}, which describes the quantum state of the battery in phase space (represented by position, $x$, and momentum, $p$). The analysis has been computed using QuTiP \cite{qutip} and the results are presented in Fig.~\ref{fig:single_qubit_wigner}a. There, we see a crescent shape symmetric in $x$ for the Wigner function around the negative value $p\approx -5$. Moreover, by comparing panels (i) and (ii) in Fig. \ref{fig:single_qubit_wigner}a we see that a bump of highly excited Fock states appears for $\lambda=10$, which is a diagnostic of the instability of the (metastable) steady state.

It can be verified that the Wigner function of a state charged according to the fully coherent charging protocol looks essentially like that of a squeezed coherent state. By decreasing the coherency parameter $c$, the Wigner function becomes more warped as we see in our plots. A similar shape can be seen for nonlinear coherent states driven by a Kerr non-linearity \cite{Roman-Ancheyta:14,Chatterjee:16}, which suggests that the dynamics of the model can be reinterpreted as a cavity interacting with a bath of qubits.

Finally, it is worth mentioning that the generic behavior described in this Section is also observed when repeating the calculation with different non fine-tuned values of $g$. Moreover, this scheme is robust, as a very similar time evolution is seen even when random noise is introduced to the optimized $(c,q)$ parameters.

\subsection{Stream of different qubit batches of identical states within}
\label{sec:results_2_batches}

Here we conduct the  more interesting optimization of the micromaser charging process described in Fig. \ref{fig:General-model}c, using a stream of $n=1000$ qubits divided into two batches consisting of $b=500$ identical qubits in each batch. The coupling constant is set to the same non fine-tuned value $g=\pi/\sqrt{15.6} \approx 0.795$ as in the previous case of Section \ref{sec:results_identical}. 

Contrary to the previous case, we now have two pairs of parameters corresponding to the first and the second batches of identical qubits: $\{\Vec{p}_1, \Vec{p_2}\} = \{(c,q)_{\rm{Batch}=1},(c,q)_{\rm{Batch}=2}\}$, respectively, which means that in the loss function, defined in Eq.~\eqref{eq:loss_function_single}, we modify $c$ and $q$ by their mean value, as described previously.
In this case we take $\eta$ to be the last $10\%$ qubits instead of the $20\%$ we took when all qubits where identical to allow a comparatively similar relaxation time between the two cases. As before, we vary the hyperparameter $\lambda$ and run the optimization algorithm several times.

We find the optimized pairs of parameters for different values of $\lambda$ to be: for $\lambda = 1$, $\{(0.150,0.079)$, $(0.361, 0.150)\}$; for $\lambda = 10$, $\{(0.229,0.123)$, $(0.414, 0.165)\}$; for $\lambda = 100$, $\{(0.270, 0.152)$, $(0.486, 0.179)\}$. It is interesting to notice that as $\lambda$ gets smaller, the optimized strategy is to first charge according to the \textit{incoherent} charging protocol followed by a new set of parameters that are able to stabilize the trajectory to a state that is long-lived while maximizing $\mathcal{W}$ as much as possible. 

Using the optimized parameters $\{\Vec{p}_1, \Vec{p_2}\}$, we simulate the time evolution described by Eq.~\eqref{eq:U_I} and calculate the ergotropy, $\mathcal{W}$ as explained in Eq.~\eqref{eq:ergotropy_definition}, for up to $k=10^5$ collisions. To make clear this point, the first set of 500 interactions are simulated using the pair of parameters corresponding to $\Vec{p}_1=(c,q)_{\rm{Batch}=1}$, whereas for the qubit $k=501$ up to the $10^5$, the parameters correspond to the second optimized pair, $\Vec{p}_2=(c,q)_{\rm{Batch}=2}$. The results are shown in Fig. \ref{fig:single_qubit}b(i). 

The main result of this first procedure are the red (circle), cyan (square) and magenta (up-triangle) lines in Fig.~\ref{fig:single_qubit}(i). Let's first focus on the red (circle) one. From this curve we see that the algorithm has been able to automatically find a transition to the second chamber, albeit the state is not stable past after $10^4$ interactions. However, the fact that this is possible is already extremely valuable. 

Motivated by the results of the previous section, the next natural step is increasing $\lambda$, whose effect can be seen in the cyan (square) and magenta (up-triangle) curves. What we see is that a greater stability can only be guaranteed by staying in the first chamber, which in turn suggests that by adding more qubits one should be able to stabilize much better well inside the second chamber. 

These observations lead naturally to what we call the \textit{improved} strategy, which makes use of two ingredients: the first one consists in fixing the parameters in the first batch to the \textit{incoherent} charging protocol regime $(c_1,q_1) = (0,0)$, such that the charging, during this first batch, becomes \textit{transparent} to the presence of the chambers. The second ingredient is increasing the number of stabilizer qubits from $500$ to $1000$ and choosing a large enough value of $\lambda$ to guarantee stability. Interestingly, we find that stability is obtained by a fairly small value of $\lambda$, i.e. $\lambda=10$. The result of this optimization is the green (cross) line in Fig.~\ref{fig:single_qubit}b, where the second pair of optimized parameters are $(c_2,q_2) = (0.449,0.208)$ and that performs notably better than the previous cases. In panel (ii) of the same Figure~\ref{fig:single_qubit}b we show the trajectories for the purity of each of the states shown in panel (i), and we observe that the purity decreases as ergotropy increases.

The second interesting thing that we notice from Fig.~\ref{fig:single_qubit}b is that by using this same strategy of first driving the instability followed by new stabilizing parameters allows to reach much more stable states in the high-end of the first chamber than with identical single copies of qubits. Again, we have checked that all the trajectories shown in Fig.~\ref{fig:single_qubit}b are stable even in the presence of noise in the $(c,q)$ parameters.

As in the previous Section, we further characterize the charged battery steady state by computing the distribution of the populations in the final state and its associated Wigner function. The results are presented in Fig. \ref{fig:single_qubit_wigner}b. In addition to the crescent shape symmetric in $x$ for the Wigner function around the negative value $p\approx -5$, another one appears around $p\approx -10$ when considering the state from the second chamber. However, there is no appreciable sign of interference between the two, suggesting that they participate as a semiclassical mixture more than a superposition. Moreover, by comparing panels (i) and (ii) within Fig.~\ref{fig:single_qubit_wigner}b, we see that the bump of highly excited Fock states is much suppressed in this new case, signaling the larger stability of this state.

\section{Conclusions and outlook}
\label{sec:conclusions}

In this paper, we have successfully developed a computational framework for optimizing parameters in quantum battery models using gradient descent. The method is very general and, as a proof of concept, we have studied the optimization of the micromaser QB in a comprehensive and detailed way, by studying two distinct charging scenarios. 
As a result, and to confirm the power of this method, we discovered a charging protocol that allows for automatic and controlled transitions between different trapping chambers in Hilbert space, ultimately leading to the stabilization of the micromaser QB and improved charging efficiency.

The potential implications and future directions prompted by this work are multiple. The novel computational framework presented in this study is both robust and highly versatile, with the potential for application to a wide range of QB models. Furthermore, it allows for large-scale optimization, thus enabling the optimization of a significant number of parameters. An immediate case of interest to consider is extending the micromaser QB charging optimization to a setup in which dissipation and other imperfections are present \cite{shaghaghi_lossy_2022, konar2022quantum}. Similarly, it would be interesting to test the effectiveness of this method by taking loss functions tailored towards figures of merit other than the ergotropy, like the charging power.
Remaining with possible further developments involving the micromaser setup, it would be important to study how to include the effect of the counter-rotating terms in the optimization protocol. In this paper, we made use of the results of \cite{huang_ultrastrong_2020} to infer that the counter-rotating terms can be ignored in the ultra-strong coupling limit, but it would be very interesting (and also relevant for potential experimental realizations) to consider their effect too. Perhaps, developments along this line may be considered by means of the well-known Magnus expansion.
Additionally, the authors of \cite{morrone2022charging}, for the particular architecture considered in that case, have been able to obtain an analytical relation between the ergotropy, the energy, and the purity of the battery's steady state. For the case of micromaser the situation is more complicated, especially since the battery states are, at best, \textit{almost} steady states. Nevertheless, it would be very interesting if such a relation could be obtained.

Additional research opportunities that stem from our findings are to explore optimized dynamics of QB models for many-body systems \cite{rossini_many-body_2019, ferraro_high-power_2018}, such as SYK \cite{rossini_quantum_2020} or spin chains \cite{le_spin-chain_2018}, which would be of particular interest from the point of view of achieving a quantum charging advantage. Another potential direction for future research would be to investigate the design of practical QB devices with the support of artificial intelligence tools. Specifically, the optimization framework hereby presented allows for the optimization of parameters within the constraints of QB model-dependent variables, which are closely aligned with real-world application requirements \cite{dou_superconducting_2022}.
Also, from a more theoretical perspective, it will be interesting to compare this approach with the reinforcement learning methods of \cite{erdman_reinforcement_2022}. Such a close comparison could show in which cases one of the two approaches is preferable.

In conclusion, this avenue of research holds great promise for the advancement of QB technology and its potential for future developments and practical applications in the field.

\section*{Acknowledgments}

CR and JO thank Vittorio Peano for useful discussions on Wigner functions. DR thanks Giuliano Benenti, Matteo Carrega, Vahid Shaghaghi and Varinder Singh for discussions and collaboration on micromaser quantum batteries. 
We thank Marcello Andolina for comments on a first version of our draft.
CR acknowledges the financial support of Spanish MINECO for the grant PRE2019-087613. JO's work is supported by the Munich Quantum Valley, which is supported by the Bavarian state government with funds from the Hightech Agenda Bayern Plus. DR acknowledges the support by the Institute for Basic Science in Korea (IBS-R024-D1).   
\bibliography{main}

\end{document}